\documentclass[10pt,twocolumn,letterpaper]{article}

\usepackage{cvpr}
\usepackage{amsmath}
\usepackage{amssymb}
\usepackage{epsfig}
\usepackage{graphicx}

\usepackage{algorithm}
\usepackage{algorithmic}
\usepackage{amsfonts}
\usepackage{caption}
\usepackage{color}
\usepackage{multirow}
\usepackage{tablefootnote}
\usepackage{times}

\usepackage[breaklinks=true,bookmarks=false]{hyperref}

\cvprfinalcopy 

\def\cvprPaperID{****} 

\begin{document}

\title{ Residual Channel Attention Generative Adversarial Network for Image Super-Resolution and Noise Reduction }

\author{\parbox{16cm}{\centering
    {\large Jie Cai, Zibo Meng, and Chiu Man Ho}\\
    {\large InnoPeak Technology, Palo Alto, CA, USA}\\
    {\normalsize \{jie.cai,\ zibo.meng,\ chiuman\}@innopeaktech.com }\\
    }
}

\maketitle
\thispagestyle{empty}
\pagestyle{empty}

\begin{abstract}
Image super-resolution is one of the important computer vision techniques aiming to reconstruct high-resolution images from corresponding low-resolution ones.
Most recently, deep learning based approaches have been demonstrated for image super-resolution.
However, as the deep networks go deeper, they become more difficult to train and more difficult to restore the finer texture details, especially under real-world settings. 
In this paper, we propose a Residual Channel Attention-Generative Adversarial Network (RCA-GAN) to solve these problems.
Specifically, a novel residual channel attention block is proposed to form RCA-GAN, which consists of a set of residual blocks with shortcut connections, and a channel attention mechanism to model the interdependence and interaction of the feature representations among different channels.
Besides, a generative adversarial network (GAN) is employed to further produce realistic and highly detailed results.
Benefiting from these improvements, the proposed RCA-GAN yields consistently better visual quality with more detailed and natural textures than baseline models; and achieves comparable or better performance compared with the state-of-the-art methods for real-world image super-resolution.
\end{abstract}

\section{INTRODUCTION}
Image super-resolution~\cite{wang2020deep,yang2019deep}, which aims to restore high-resolution images from corresponding low-resolution ones, is an important class of image processing and computer vision techniques. 
Image super-resolution is widely used in a range of real-world applications, such as medical image~\cite{huang2017simultaneous}, surveillance~\cite{zhang2010super} and oceanography~\cite{ducournau2016deep}. 
However, image super-resolution is a highly challenging and inherently ill-posed problem since there always exists multiple possible high-resolution solutions corresponding to a low-resolution image.

With the rapid development of deep learning techniques~\cite{cai2019improving,cai2018island,cai2019feature,li2019Pooling} recently, deep learning based image super-resolution models~\cite{dong2014learning,kim2016accurate,kim2016deeply,ledig2017photo,lim2017enhanced,sajjadi2017enhancenet,tai2017image,tai2017memnet,tong2017image,zhang2019residual} have been actively developed and achieved the notable performance.
Specifically, the achievement starts from the promising super-resolution methods using Convolutional Neural Networks (CNNs) (e.g., SRCNN~\cite{dong2014learning}) to recent Generative Adversarial Nets (GANs)~\cite{goodfellow2014generative} based approaches (e.g., SRGAN~\cite{ledig2017photo}).
In general, the design of deep learning based super-resolution methods differ from each other in the following major aspects: (1) different types of network structures; (2) different types of loss functions; (3) different types of training strategies and principles.
However, simply stacking trainable network layers to construct deeper networks can hardly obtain better performance in terms of peak signal-to-noise ratio (PSNR) and perceptual quality.
Therefor, how to construct very deep networks and whether deeper networks can further boost image super-resolution performance still remains a challenge.

In this work, we increase network depth by utilizing Residual Net (ResNet)~\cite{he2016deep}, which makes it possible to train up to hundreds of layers without suffering from gradient vanishing problem and still achieves compelling performance. 
Besides, most recent CNN-based methods~\cite{dong2014learning,kim2016accurate,kim2016deeply,tai2017memnet,zhang2018residual} lack ability in dealing with different types of information across feature channels and hence hinder the representational power of deep networks.
To solve these problems, we propose a residual channel attention block to build up a very deep trainable network and adaptively rescale each channel-wise feature by modeling the interdependencies across feature channels simultaneously. Such channel attention mechanism allows our proposed network to concentrate on more useful channels and enhance discriminative learning ability. 
Generally speaking, the low-resolution images contain more low-frequency information, which can directly forwarded to the final high-resolution outputs. 
Therefore, to ease the training of very deep networks, we introduce residual in residual structure, where the shortcut connection located in the stacked residual channel attention blocks and the long skip connection allow abundant low-frequency information to be directly bypassed through the identity-based skip connection, which could ease the flow of information. 
Furthermore, the proposed RCA-GAN is capable of generating nature textures and fine details by cooperating GAN technique.

In summary, our major contributions are:

\begin{enumerate}
\item[-] Proposing a very deep Residual Channel Attention-Generative Adversarial Network (RCA-GAN) for real-world image super-resolution.
\item[-] Developing channel attention mechanism to adaptively rescale feature channels by considering interdependencies among feature channels and selectively emphasizing informative features.
\item[-] Employing generative adversarial network to generate more visually pleasing and nature results.
\end{enumerate}

Extensive experiments show that the proposed RCA-GAN yields better perceptual quality and visual super-resolution results against the baseline models; and achieves comparable or better performance compared with the state-of-the-art methods for real-world image super-resolution.

\section{RELATED WORK}
\begin{figure*}[th!]
   \centering
   \includegraphics[width=0.9\textwidth]{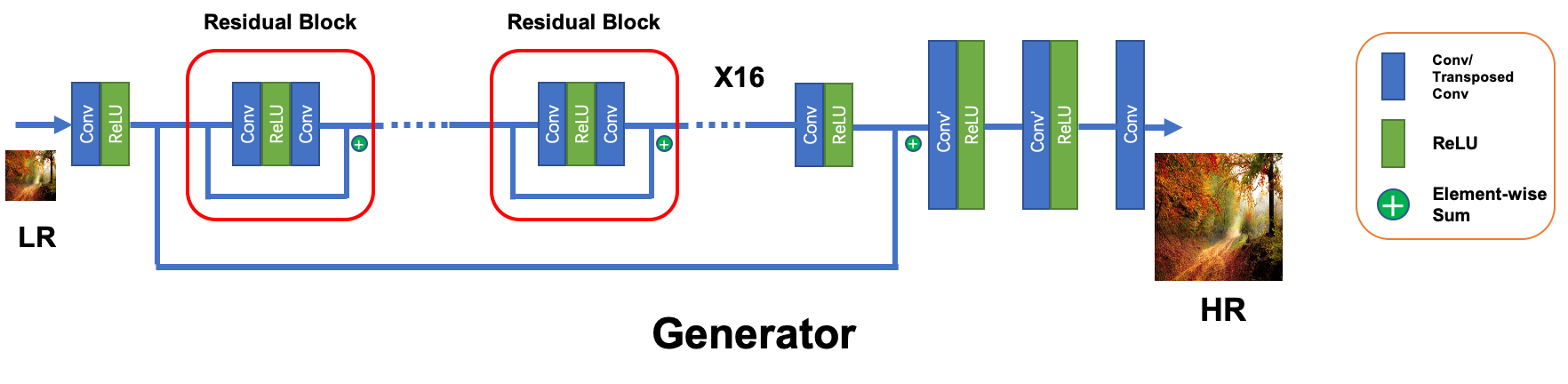}
   \caption{\small{An illustration of the baseline Residual Network (RN). The RN consists of 16 recursive residual blocks. Best viewed in color.}}
   \label{fig:G1}
\end{figure*}

Numerous image super-resolution methods have been studied in the image processing and computer vision field. Recently, the powerful capability of deep CNNs has led to dramatic improvements in image super-resolution. Moreover, attention mechanism and generative adversarial network have been fairly popular concepts and useful techniques in the deep learning community. In this section, we focus on works related to CNN-based methods, GAN-based methods, and attention mechanism for image super-resolution. Then we introduce the concept of real-world image super-resolution.

\textbf{Deep CNN for image super-resolution.}
The pioneer work was achieved by Dong et al.~\cite{dong2014learning}, who developed SRCNN by introducing a three-layer CNN for image super-resolution and achieved superior performance than previous work. 
Kim et al. increased the CNN depth to 20 in VDSR~\cite{kim2016accurate} and DRCN~\cite{kim2016deeply} and achieved notable improvements over SRCNN later on.
He at al.~\cite{he2016deep} proposed ResNet which could increase the network depth up to hundreds of layers without hurting the performance.
Such effective residual learning strategy was then introduced in many other CNN-based image super-resolution methods~\cite{ledig2017photo,lim2017enhanced,sajjadi2017enhancenet,tai2017image,tai2017memnet,tong2017image,zhang2018residual}. 
Lim et al.~\cite{lim2017enhanced} proposed a very deep network MDSR and a very wide network EDSR by utilizing multiple residual blocks.
Tai et al.~\cite{tai2017image} developed a very deep yet concise Deep Recursive Residual Network (DRRN) which consists up to 52 convolutional layers.
Specifically, residual learning is deployed both in local and global manners to ease the difficulty of training very deep networks.
Huang et al.~\cite{huang2017densely} proposed Dense Convolutional Network (DenseNet), which connects each layer to every other layer in a feed-forward fashion. 
DenseNet strengthens feature propagation, encourages feature reuse, and substantially reduces the number of parameters.
Tong et al.~\cite{tong2017image} and Zhang et al.~\cite{zhang2018residual} jointly deployed residual block and dense block in very deep networks, which provide an effective way to combine the low-level features and high-level features and further boost the reconstruction performance for image super-resolution.

\textbf{Deep GAN for image super-resolution.}
Most recently, the GANs~\cite{goodfellow2014generative} receive more and more attention and are introduced to various computer vision tasks due to the powerful learning ability. 
It is straightforward to employ adversarial learning in image super-resolution, in which case we only need to treat the super-resolution model as the generator, and define an extra discriminator to judge whether the input image is from generator or real data. 
Therefore, Ledig et al.~\cite{ledig2017photo} firstly proposed SRGAN using adversarial loss and achieved compelling performance.
Besides, the ESRGAN~\cite{wang2018esrgan} employed relativistic GAN~\cite{jolicoeur2018relativistic} to train the generator which not only increase the probability that fake data is real but also decrease the probability that real data is real. 
In other word, the discriminator predicts the probability that how real images are relatively more realistic than fake ones, instead of the probability that input images are just real or fake, and thus leads to recover more detailed textures.

\textbf{Attention Mechanism for image super-resolution.}
Attention mechanism has improved the success of various computer vision tasks recently and continues to be an omnipresent component in state-of-the-art models.
In broad terms, attention can be viewed as a guidance to bias the allocation of available processing resources towards the most informative components of an input.
Considering the interdependence and interaction of the feature representations between different channels, Hu et al.~\cite{hu2018squeeze} proposed a squeeze-and-excitation block to improve learning ability by explicitly modeling channel interdependence. 
Recently, Zhang et al.~\cite{zhang2018image} incorporated the channel attention mechanism with super-resolution and proposed RCAN, which improves the representation ability of the model and super-resolution performance. 
In order to better learn the feature correlations, Dai et al.~\cite{dai2019second} further proposed a second-order channel attention (SOCA) module. The SOCA adaptively rescales the channel-wise features by using second-order feature statistics instead of GAP, and enables extracting more discriminative and informative representations.

Most existing super-resolution models have very limited local receptive fields. However, some distant objects or textures may be very important for local patch generation. 
So that Zhang et al.~\cite{zhang2019residual} proposed local and nonlocal attention blocks to extract features that capture the long-range dependencies between pixels.
Through this mechanism, the proposed method captures the spatial attention well and further enhances the representation ability. Similarly, Dai et al.~\cite{dai2019second} also incorporated the non-local attention mechanism to capture long-distance spatial contextual representations for image super-resolution.

\textbf{Real-world image super-resolution.}
Most recent proposed approaches rely on paired low-resolution and high-resolution images to train the deep network in a fully supervised manner. However, such image pairs are not often available in most real-world applications. 
The AIM 2019 Challenge on Real-World Image Super-Resolution~\cite{AIM2019RWSRchallenge} and the NTIRE 2020 Challenge on Real-World Image Super-Resolution~\cite{NTIRE2020RWSRchallenge} aim to stimulate research in the direction of real-world image super-resolution, i.e., no paired reference high-resolution images are provided for training. 
Fritsche et al.~\cite{fritsche2019frequency} proposed DSGAN which trained in an unsupervised manner on high-resolution images. Specifically, they first generated low-resolution images with the same characteristics as the original images; and then utilized the generated data to train a super-resolution model, which improves the performance on real-world images. 
Lugmayr et al.~\cite{lugmayrICCVW2019} proposed a super-resolution model which can be trained under the supervision of direct pixel-wise in the high resolution domain, while robustly generalizing to real input.

\section{METHODOLOGY}
\begin{figure*}[th!]
   \centering
   \includegraphics[width=0.86\textwidth]{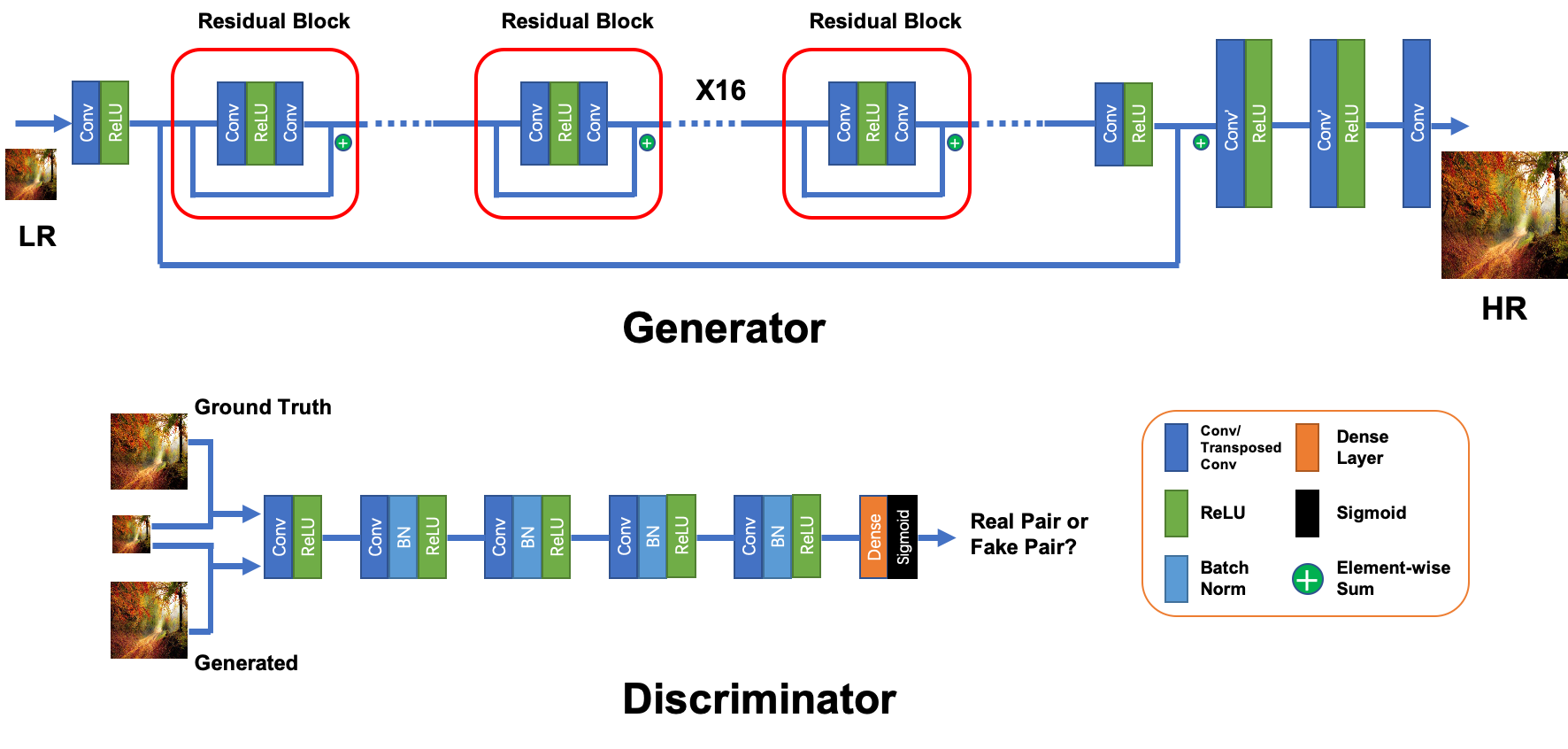}
   \caption{\small{An illustration of the baseline Residual Network-Generative Adversarial Network (RN-GAN). The generator of the RN-GAN consists of 16 recursive residual blocks. The discriminator of the RN-GAN consists of 10 trainable layers and the final Sigmoid function is used to produce probability. Best viewed in color.}}
   \label{fig:G2}
\end{figure*}

In this section, we will first introduce the two baseline models and the proposed RCA-GAN. 
Then we will present the proposed residual channel attention block.
Finally, we will show the overall loss function of the RCA-GAN.

\subsection{An overview of the two baseline models and proposed RCA-GAN}

We construct our first baseline Residual Network (RN) with 16 recursive residual blocks as shown in Fig.~\ref{fig:G1}. 
The residual network consists four parts: shallow feature extraction, residual block, upscale module, and reconstruction part. 
We only use one $3\times3$ convolution layer (pad 1, stride 1, and channel 128) followed by one ReLU layer to extract the shallow features.
We deploy 16 recursive residual blocks to extract the mid-level and high-level features.
The residual block consists of two $3\times3$ convolution layer (pad 1, stride 1, and channel 128) and only the first convolution layer followed by a ReLU layer.
Following the previous work~\cite{wang2018esrgan}, we also remove Batch Normalization (BN)~\cite{ioffe2015batch} layers which could scale information of each image, get rid of range flexibility from networks and lead to decrease the performance substantially.
In this work, we use two repeated $3\times3$ transposed convolution layer (stride 2 and channel 128) as upscale module.
At the final convolution layer, a $3\times3$ convolution layer (pad 1, stride 1, and channel 3) is applied to recover high-resolution outputs.

As we mentioned earlier, the GANs~\cite{goodfellow2014generative} are introduced to various computer vision tasks and yield consistently better performance. GANs are deep neural architectures used to generate images with two types of networks involved: a generator ($\mathbf{\emph{G}}$) and a discriminator ($\mathbf{\emph{D}}$). Specifically, a generator is trained to capture the underlying distribution of the training data, while a discriminator is trained to differentiate whether a sample comes from the real distribution or from the generator. The objective of a GAN is to train a $\mathbf{\emph{D}}$ that identifies fake samples generated by $\mathbf{\emph{G}}$ from samples drawn from the true distribution, while encouraging $\mathbf{\emph{G}}$ to generate realistic samples to deceive $\mathbf{\emph{D}}$. In contrast to traditional GANs that learn a mapping from the random noise vector $\mathbf{z}$ to a target sample $\mathbf{y}$, i.e., $\mathbf{\emph{G}}(\mathbf{z}) \rightarrow \mathbf{y}$, conditional GANs (cGANs)~\cite{mirza2014conditional} learn a mapping from a random noise vector $\mathbf{z}$ to the target $\mathbf{y}$ conditioned on an observed signal $\mathbf{x}$, i.e., $\mathbf{\emph{G}}(\mathbf{x},\mathbf{z}) \rightarrow \mathbf{y}$. In this work, cGAN is utilized to perform image super-resolution task.

As shown in Fig.~\ref{fig:G2}, the second baseline Residual Network-Generative Adversarial Network (RN-GAN) consists two networks: the generator ($\mathbf{\emph{G}}$), i.e., Residual Network (RN) and discriminator ($\mathbf{\emph{D}}$). The proposed discriminator consists of one $3\times3$ convolution layers (pad 1, stride 1, and channel 64) followed by a leaky ReLU (slope 0.2), two repeated implementations of $4\times4$ convolution layers (pad 1, stride 2, and channel 64) followed by a Batch Norm (BN) layer and a leaky ReLU (slope 0.2), another two repeated implementation of $4\times4$ convolution layers (pad 1 and stride 2) followed by a BN layer and a LReLU layer. At the last two convolution layers, we double the number of output feature channels. At the final convolution layer, a dense layer is applied to generate 1-dimensional output, and then followed by a Sigmoid function.

\begin{figure*}[th!]
   \centering
   \includegraphics[width=0.86\textwidth]{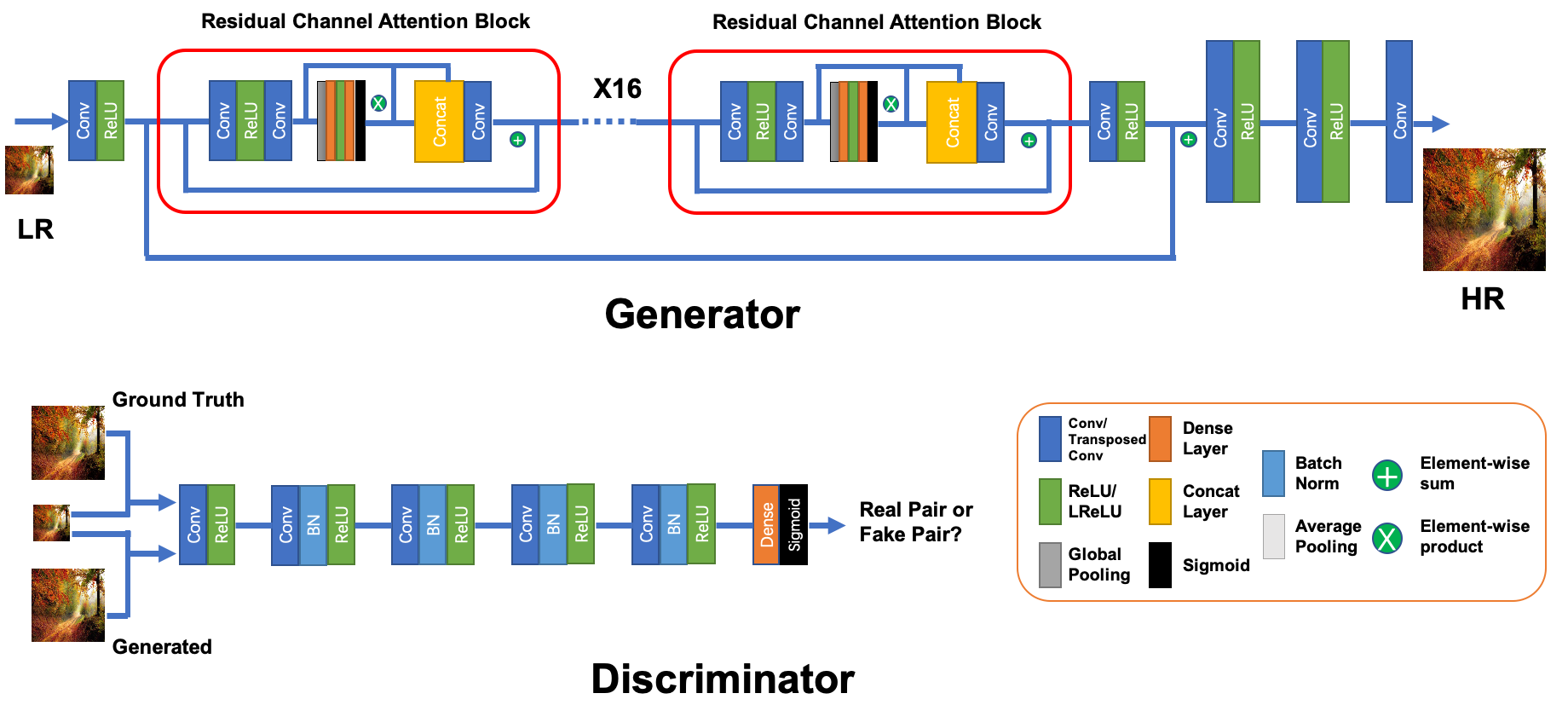}
   \caption{\small{An illustration of the proposed Residual Channel Attention-GAN (RCA-GAN). The generator consists of 16 residual channel attention blocks. Best viewed in color.}}
   \label{fig:G3}
\end{figure*}

\begin{figure*}[th!]
   \centering
   \includegraphics[width=0.88\textwidth]{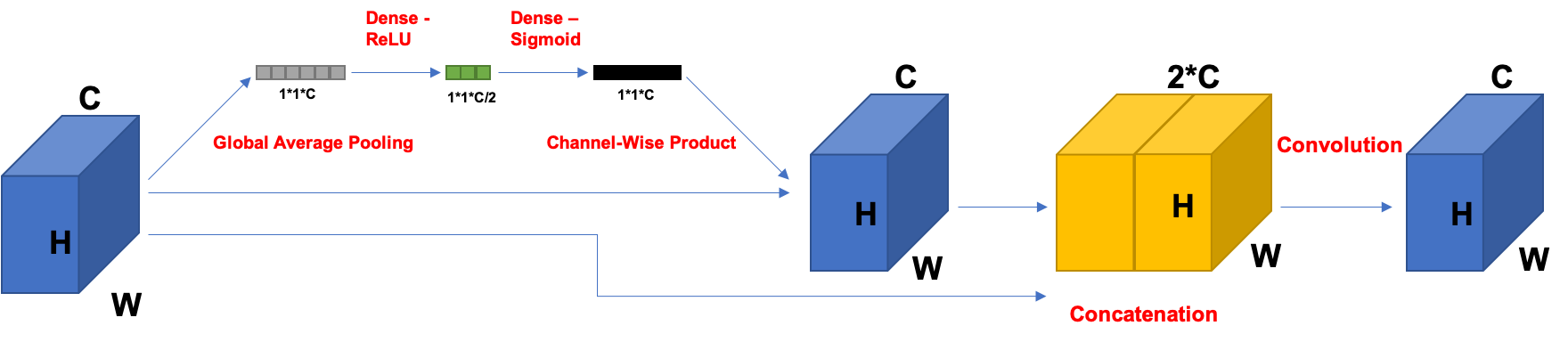}
   \caption{\small{An illustration of the proposed channel attention module. Best viewed in color.}}
   \label{fig:CAM}
\end{figure*}

As shown in Fig.~\ref{fig:G3}, the proposed Residual Channel Attention-GAN (RCA-GAN) consists two networks: residual channel attention generator ($\mathbf{\emph{G}}$) and discriminator ($\mathbf{\emph{D}}$). The discriminator ($\mathbf{\emph{D}}$) is identical to the second baseline model. The residual channel attention generator also consists four parts: shallow feature extraction, residual channel attention block, upscale module, and reconstruction part. 
The shallow feature extraction part, upscale module, and reconstruction part are identical to the Residual Network (RN).
Previous CNN-based super-resolution methods don't explicitly model the relationship between different channels, which is not flexible for the real-world applications and lack learning ability to capture global information. 
In order to make the network focus on more informative features, we exploit the interdependencies among different feature channels and provide a way of modulating the channel activation, resulting in a channel attention mechanism as shown in Fig.~\ref{fig:CAM}.

\subsection{A Channel Attention Module}

In the proposed channel attention module (Fig.~\ref{fig:CAM}), each input channel is squeezed into a channel descriptor, i.e., a constant, using global average pooling (GAP), then these descriptors are fed into two dense layers (followed by a ReLU layer and a Sigmoid layer, respectively) to produce channel-wise scaling factors for input channel. The input channel and the scaled input channel are concatenated across channels and then feed into a $3\times3$ convolution (pad 1, stride 1, and channel 128) layer. 
The proposed channel attention module adaptively recalibrates channel-wise feature responses by explicitly modeling interdependencies between different channels. 
The proposed channel attention block achieves better performance than traditional channel attention block~\cite{hu2018squeeze,zhang2018image} by deploying the concatenation and convolution operations which could better capture both spatial and channel dependence.
Note that, the output channel retains the same dimension as the input channel. Therefore, our proposed residual channel attention module is very flexible and can be easily applied to any CNN models.

\subsection{Loss Functions of the RCA-GAN}

The overall loss function of the RCA-GAN is defined as:
\begin{equation} \label{eq:joint_IF-GAN}
 \begin{aligned}
& \mathcal{L} = \lambda_{1} \cdot \mathcal{L}_{cGAN}(\mathbf{\emph{G}},\mathbf{\emph{D}}) + \lambda_{2} \cdot \mathcal{L}_{L1}(\mathbf{\emph{G}}) + \lambda_{3} \cdot \mathcal{L}_{Gradient}(\mathbf{\emph{G}}) 
\\ &
+ \lambda_{4} \cdot \mathcal{L}_{VGG} (\mathbf{\emph{G}})  + \lambda_{5} \cdot \mathcal{L}_{SSIM} (\mathbf{\emph{G}}) + \lambda_{6} \cdot \mathcal{L}_{multi\_SSIM}(\mathbf{\emph{G}}) 
 \end{aligned}
\end{equation}
where the hyperparameters $\lambda_{1}=1$, $\lambda_{2}=1$, $\lambda_{3}=10$, $\lambda_{4}=0.2$, $\lambda_{5}=0.1$, and $\lambda_{6}=0.1$ are used to balance the six terms.

\subsubsection{Adversarial Loss}
The first term of Eq.~\ref{eq:joint_IF-GAN} is the loss function of a cGAN and defined as:
\begin{equation} \label{eq:cGAN1}
 \begin{aligned}
\mathcal{L}_{cGAN}(\mathbf{\emph{G}},\mathbf{\emph{D}}) & =  {\mathbb{E}} [log(\mathbf{\emph{D}}(\{ \mathbf{I}_{_{LR}}, \mathbf{I}_{_{HR}}\}))]
\\ &
+ {\mathbb{E}} [log(1-\mathbf{\emph{D}} ( \{ \mathbf{I}_{_{LR}}, \mathbf{\emph{G}}(\mathbf{I}_{_{LR}})\})  )]
 \end{aligned}
\end{equation}
where $\mathbf{\emph{G}}(\mathbf{I}_{_{LR}})$ represents the generated high-resolution image, $\{\mathbf{I}_{_{LR}}, \mathbf{\emph{G}}(\mathbf{I}_{_{LR}})\}$ denotes the fake tuple, and $\{\mathbf{I}_{_{LR}}, \mathbf{I}_{_{HR}}\}$ denotes the real tuple.

\subsubsection{Pixel Loss}
To compete against the discriminator $\mathbf{\emph{D}} $, $\mathbf{\emph{G}}(\cdot)$ learns to capture the true data distribution to generate realistic images that are similar to the images sampled from the true data distribution. We explore this option using pixel-wise $L1$ loss between two images:
\begin{equation} \label{eq:L1}
\mathcal{L}_{L1}(\mathbf{\emph{G}})={\mathbb{E}} [\lVert \mathbf{I}_{_{HR}} -\mathbf{\emph{G}}( \mathbf{I}_{_{LR}}) \rVert_{1}]
\end{equation}

The pixel-wise $L1$ loss constrains the generated $\mathbf{\emph{G}}( \mathbf{I}_{_{LR}} )$ to be close enough to the ground truth $\mathbf{I}_{_{HR}} $ on the pixel values.
Comparing with $L_1$ loss, the $L_2$ loss penalizes larger errors but is more tolerant to small errors, and thus often results in too smooth results. In practice, the $L_1$ loss shows better performance and convergence over $L_2$ loss.
Since the definition of PSNR is highly correlated with pixel-wise difference thus minimizing pixel loss directly maximize PSNR, the pixel loss gradual becomes the most widely used loss function.

In addition to the $\mathcal{L}_{L1}(\mathbf{\emph{G}})$, we also implement pixel-wise $L1$ loss on image gradients in both vertical and horizontal directions:

\begin{equation} \label{eq:gradient}
 \begin{aligned}
\mathcal{L}_{Gradient}(\mathbf{\emph{G}}) & = 0.5 \cdot {\mathbb{E}} [\lVert {\mathbf{I}_{_{HR}}}_{x} - {\mathbf{\emph{G}}( \mathbf{I}_{_{LR}})}_{x} \rVert_{1}] 
\\ &
+ 0.5 \cdot {\mathbb{E}} [\lVert {\mathbf{I}_{_{HR}}}_{y} - {\mathbf{\emph{G}}( \mathbf{I}_{_{LR}})}_{y} \rVert_{1}]
 \end{aligned}
\end{equation}
where $ {\mathbf{I}_{_{HR}}}_{x} $ and $ {\mathbf{I}_{_{HR}}}_{y} $ represent the ground truth high-resolution image gradient in horizontal and vertical directions, respectively. $ {\mathbf{\emph{G}}( \mathbf{I}_{_{LR}})}_{x} $ and $ {\mathbf{\emph{G}}( \mathbf{I}_{_{LR}})}_{y} $ represent the generated high-resolution image gradient in horizontal and vertical directions, respectively.

\subsubsection{Content Loss}
However, since the pixel loss actually doesn’t take image quality (e.g., perceptual quality~\cite{johnson2016perceptual}, image texture~\cite{sajjadi2017enhancenet}) into consideration, the outputs often lack high-frequency details and are perceptually unsatisfying with oversmooth textures.
In order to further improve perceptual quality of images, the content loss is introduced into super-resolution~\cite{johnson2016perceptual,dosovitskiy2016generating}. 
Specifically, it measures the semantic differences between images using a pre-trained image classification network, e.g., VGG-19. 
Denoting this network as $\phi$ and the extracted high-level representations on $l$-th layer as $\phi^{l}(I)$, the content loss is indicated as the $L1$ distance and cosine distance between high-level representations of two images:

\begin{equation}  \label{eq:VGG}
 \begin{aligned}
\mathcal{L}_{VGG} (\mathbf{\emph{G}}) & = \lambda \cdot  {\mathbb{E}} [\lVert \phi^{l}({\mathbf{I}_{_{HR}}}) - \phi^{l}({\mathbf{\emph{G}}( \mathbf{I}_{_{LR}})}) \rVert_{1}]
\\ &
+ {\mathbb{E}} [ 1 - \phi^{l}({\mathbf{I}_{_{HR}}}) \cdot \phi^{l}({\mathbf{\emph{G}}( \mathbf{I}_{_{LR}})}) ]
 \end{aligned}
\end{equation}

Note that, $\phi^{l}({\mathbf{I}_{_{HR}}}) \cdot \phi^{l}({\mathbf{\emph{G}}( \mathbf{I}_{_{LR}})})$ is defined as a cosine function measuring the similarity between two vectors. And the hyperparameters $\lambda=10$ is used to balance the two distances. 

Essentially the content loss transfers the learned knowledge of hierarchical image features from the trained classification network $\phi$ to the target super-resolution network. In contrast to the pixel loss, the content loss encourages the output image to be perceptually similar to the ground truth image instead of forcing them to match pixel values exactly.

\subsubsection{SSIM Loss and multi-scale SSIM Loss}
Considering that the human visual system is highly adapted to extract image structures, the structural similarity index (SSIM)~\cite{wang2004image} is proposed for measuring the structural similarity between images based on independent comparisons in terms of luminance, contrast, and structures.
Besides, the multi-scale structural similarity (MS-SSIM)~\cite{wang2003multiscale} supplies more flexibility than traditional SSIM in incorporating the variations of viewing conditions.
Therefore, we further deploy $ \mathcal{L}_{SSIM} (\mathbf{\emph{G}}) $ and $ \mathcal{L}_{multi\_SSIM}(\mathbf{\emph{G}}) $ into the proposed network to better generate images in perceptual aspect.

\section{EXPERIMENTS}
\begin{table*}
\setlength{\tabcolsep}{0.1in}
  \begin{center}
    \caption{\small{Performance comparison for the Track 1 in terms of the PSNR, SSIM, LPIPS, and MOS, respectively. }}
    \label{tab:results}
    \begin{tabular}{c|c|c|c|c}
    \hline
     Method    &     PSNR$\uparrow$     &       SSIM$\uparrow$      &      LPIPS$\downarrow$     &   MOS$\downarrow$    \\
    \hline
    Impressionism &  24.67 (16) &  0.683 (13) &  0.232 (1)        &  2.195 \\
    Samsung-SLSI-MSL &  25.59 (12) &  0.727 (9) &  0.252 (2)   &  2.425 \\
    BOE-IOT-AIBD &  26.71 (4) &  0.761 (4) &  0.280 (4)            &  2.495 \\
    MSMers &  23.20 (18) &  0.651 (17) &  0.272 (3)                   &  2.530 \\
    KU-ISPL &  26.23 (6) &  0.747 (7) &  0.327 (8)                      &  2.695 \\
    \textbf{InnoPeak-SR (RCA-GAN)} &  \textbf{26.54 (5)} &  \textbf{0.746 (8)} &  \textbf{0.302 (5)}    &  \textbf{2.740} \\
    ITS425 &  27.08 (2) &  0.779 (1) &  0.325 (6)                       &  2.770 \\
    MLP-SR &  24.87 (15) &  0.681 (14) &  0.325 (7)                  &  2.905 \\
    Webbzhou &  26.10 (9) &  0.764 (3) &  0.341 (9) &  - \\
    SR-DL &  25.67 (11) &  0.718 (10) &  0.364 (10) &  - \\
    TeamAY &  27.09 (1) &  0.773 (2) &  0.369 (11) &  - \\
    BIGFEATURE-CAMERA &  26.18 (7) &  0.750 (6) &  0.372 (12) &  - \\
    BMIPL-UNIST-YH-1 &  26.73 (3) &  0.752 (5) &  0.379 (13) &  - \\
    SVNIT1-A &  21.22 (19) &  0.576 (19) &  0.397 (14) &  - \\
    KU-ISPL2 &  25.27 (14) &  0.680 (15) &  0.460 (15) &  - \\
    SuperT &  25.79 (10) &  0.699 (12) &  0.469 (16) &  - \\
    GDUT-wp &  26.11 (8) &  0.706 (11) &  0.496 (17) &  - \\
    SVNIT1-B &  24.21 (17) &  0.617 (18) &  0.562 (18) &  - \\
    SVNIT2 &  25.39 (13) &  0.674 (16) &  0.615 (19) &  - \\
    \hline
    Bicubic &  25.48 (-) &  0.680 (-) &  0.612 (-) &  3.050 \\
    ESRGAN Supervised &  24.74 (-) &  0.695 (-) &  0.207 (-) &  2.300 \\
    \hline
	\end{tabular}
  \end{center}
\end{table*}

\begin{figure*}[th!]
   \centering
   \includegraphics[width=0.95\textwidth]{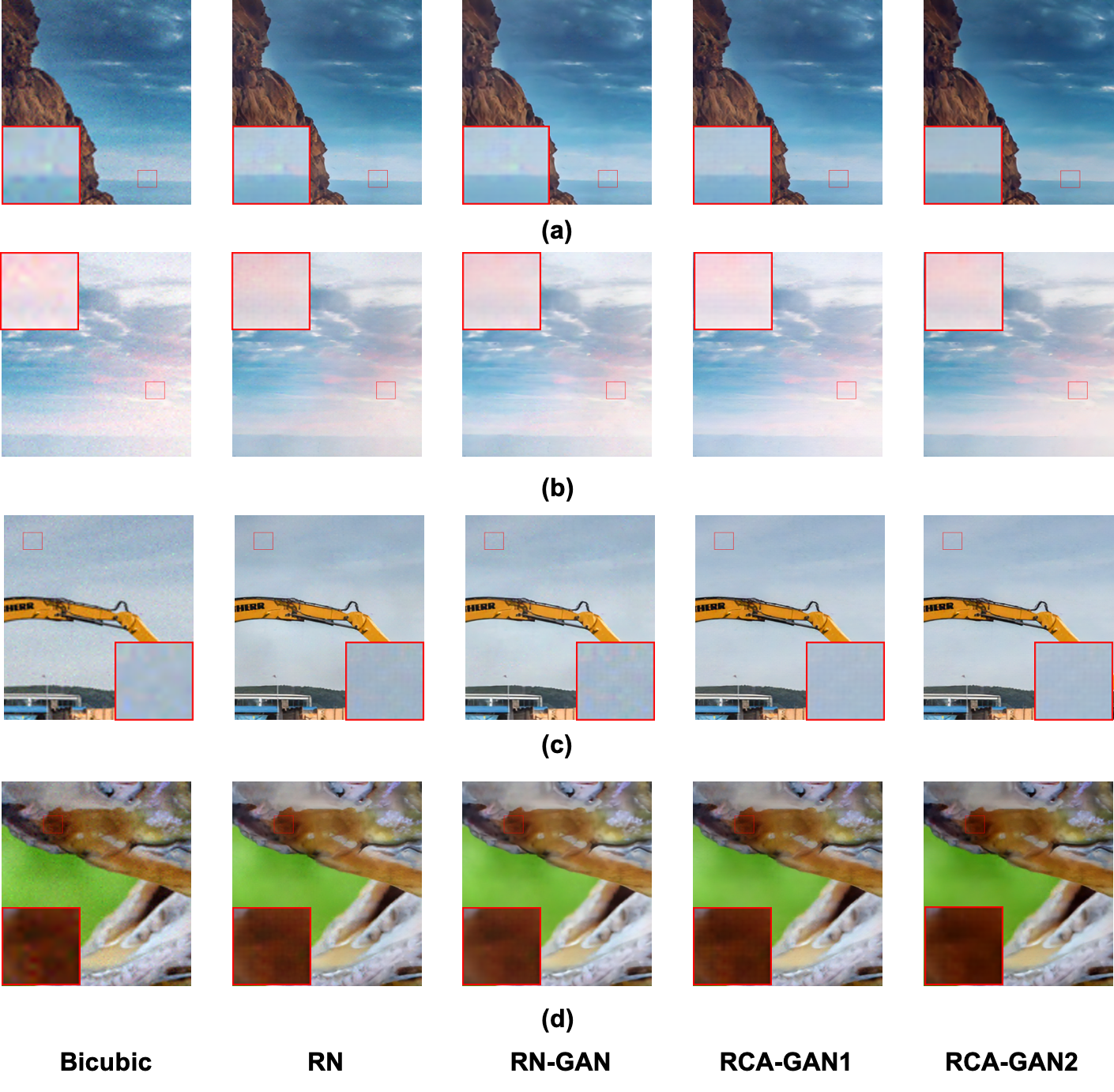}
   \caption{\small{A visual comparison for the traditional bicubic method, the baseline models (RN and RN-GAN), and the proposed RCA-GANs. Note that, the RCA-GAN1 is designed using the same structure as shown in Fig.~\ref{fig:G3}. In order to avoid checkerboard pattern, we further developed the RCA-GAN2 by deploying one more convolution layer in the reconstruction part. Best viewed in color.}}
   \label{fig:comparison}
\end{figure*}

To evaluate the proposed RCA-GAN, experiments have been conducted on two NTIRE 2020 Real World Super-Resolution Challenges~\cite{lugmayrICCVW2019,AIM2019RWSRchallenge,NTIRE2020RWSRchallenge}, i.e., Track 1: Image Processing Artifacts and Track 2: Smartphone Images.

\subsection{Experimental Datasets}
\noindent  \emph{\textbf{Track 1: Image Processing Artifacts}} is to super-resolve degradation images from the source domain to the target domain. The source domain consists of images with synthetically generated image processing artifacts and noises. Not only should those images be super-resolved with factor 4, but they should also have a clean high-quality appearance (target domain images). Specifically, this challenge contains 2,650 source domain training images, 800 target domain training images, 100 source domain validation images, and another 100 source domain testing images.

\noindent  \emph{\textbf{Track 2: Smartphone Images}} is to super-resolve image from the smart phone images to the target domain. The source domain images contain a high noise level for low light conditions and artifacts originating from low-quality smart phone images. Similar to Track 1, the Track 2 should not only super-resolve these images with factor 4, but they should also have a clean high quality appearance (target domain images). Specifically, it contains 5,902 source domain training images, 800 target domain training images same with Track 1, 113 source domain validation images, and another 100 source domain testing images.

\noindent  \emph{\textbf{Training Setting.}}  All experiments are performed with a scale factor of 4 between low-resolution images and high-resolution images. This corresponds to a 16 expansion in image pixels. In this work, we treat source domain training images and target domain training images as input high-resolution images. We aim to develop a robust architecture capable of generalizing the degradation including noise and artifacts present in the real-world setting. We obtain the low-resolution images by downsampling the high-resolution images using bicubic kernel with downsampling factor 4. Further, we collect 7,418 high-quality images from the Internet as extra high-resolution training data; and corresponding bicubic downsampling images with Gaussian noise, Poison noise, and salt $\&$ pepper noise for data augmentation purpose as low-resolution input images. During the training step, $64 \times 64$ patches are randomly cropped and horizontally flipped from the low-resolution images. The whole image patch is used for testing since the network is fully convolutional.

\subsection{Training Details}
In this work, the experiments are implemented using TensorFlow toolbox~\cite{tensorflow2015}. Adam optimizer with a mini-batch size of 50, $\beta1=0.9$, $\beta2=0.99$, and a weight decay parameter of 1e-4, is used for training the baseline models and RCA-GAN. The total number of iterations $T$ is 10,000. The learning rate $\alpha$ is 1e-4. We scale the range of the low-resolution input images to (0, 1) and for the high-resolution images to (-1, 1). The pixel loss and content loss are thus calculated on images of intensity range (-1, 1). The SSIM loss and multi-scale SSIM loss are calculated on images of intensity range (0, 1).

\subsection{Image Quality Assessment and Experimental Results}

In general, image quality assessment includes subjective methods, i.e., how realistic the image looks, and objective computational methods, e.g., PSNR, SSIM, and LPIPS.
Peak signal-to-noise ratio (PSNR) is one of the most popular reconstruction quality measurement and is the most widely used evaluation criteria for super-resolution models. 
Structural similarity index (SSIM) is proposed for measuring the structural similarity between images, based on independent comparisons in terms of luminance, contrast, and structures. 
Learned perceptual image patch similarity (LPIPS) according to the difference in deep features by trained deep networks is widely used to better assess the image perceptual quality. 
Mean opinion score (MOS) is an objective measurement obtained from a group of experts whom are asked to evaluate
the quality of the super-resolution image w.r.t. the reference image.
As show in Table~\ref{tab:results}, the proposed RCA-GAN ranks 5th, 8th, 5th, and 6th places based on PSNR, SSIM, LPIPS, and MOS measurement, respectively.
ESRGAN Supervised network is fine-tuned in a fully supervised manner by applying the synthetic degradation operation which is unknown for challenge participants. Therefore, this method serves as an upper bound in performance.

\subsection{Ablation Study}

As shown in Fig.~\ref{fig:comparison}, we show visual comparison for the bicubic method, the two baseline models (RN and RN-GAN), and the proposed RCA-GANs, respectively. Note that, the RN and RN-GAN are trained under the same learning settings with the proposed RCA-GANs. For the bicubic method, we observe that the noises and artifacts are even exacerbated since the interpolation-based upsampling methods improve the image resolution only based on its own image signal, without bringing any more information. Instead, they often introduce some side artifacts, e.g., noise amplification and blurring results. 

\textbf{Importance of Generative Adversarial Network (GAN).}
As shown in Fig.~\ref{fig:comparison} (column 2 and column 3), we observe that the RN-GAN with adversarial loss achieves more finer and realistic details than RN without adversarial loss.

\textbf{Importance of Channel Attention Mechanism.}
We can obviously find the proposed RCA-GANs can alleviate both noises and artifacts; and recover more details compared with the two baseline models. 
Such obvious comparisons demonstrate that networks with more powerful representational ability can extract more sophisticated features from the low-resolution space.
Most importantly, the two baseline models can't recover the right brightness and image pattern either. 
For example, as shown in Fig.~\ref{fig:comparison} (b) and (d), the brightness in the regions highlighted by the RCA-GANs are same with the bicubic method (bicubic doesn't change image patterns), while the two baseline models alter the image brightness since they lack global learning ability.

\textbf{Analysis on Checkerboard Pattern.}
In order to avoid the checkerboard pattern caused by the transposed convolution layer, we further employ one more convolution layer (pad 1, stride 1, and channel 128) in the final reconstruction part, i.e., the RCA-GAN2 model. 
As shown in Fig.~\ref{fig:comparison} (column 4 and column 5), we observe that the checkerboard pattern is disappeared for RCA-GAN2.

\section{CONCLUSION}
In this paper, we propose a residual channel attention generative adversarial network (RCA-GAN) for image super-resolution. 
Specifically, the residual in residual structure with shortcut skip connection and long skip connection allows abundant low-frequency information to be bypassed through identity-based skip connections, making the network mostly focus on learning high-frequency information.
Besides, to improve learning ability of the network, we propose channel attention block to adaptively rescale channel-wise features by considering interdependencies among channels. 
Moreover, we further deploy generative adversarial network to generate better visual super-resolution results.
Extensive experiments demonstrate the effectiveness of our proposed RCA-GAN.

{\small
\bibliographystyle{ieee_fullname}
\bibliography{../../../reference/abbrev_short,../../../reference/my_paper,../../../reference/my_paper_OPPO,../../../reference/Deep_Learning/deep_learning_toolbox,../../../reference/Deep_Learning/deep_learning_technique,../../../reference/Deep_Learning/CNN,../../../reference/Deep_Learning/GAN,../../../reference/Image_Super_Resolution/NTIRE2020,../../../reference/Image_Super_Resolution/Image_SR_papers,../../../reference/Image_Super_Resolution/Image_SR_application,../../../reference/Image_Super_Resolution/Image_SR_measurement}
}

\end{document}